# Electro-optically tunable low phase-noise microwave synthesizer in an active lithium niobate microdisk

Renhong Gao,[1,5] Botao Fu,[1,6] Ni Yao,[2] Jianglin Guan,[3,4] Haisu Zhang,[3] Jintian Lin,[1,5,†] Chuntao Li,[3,4] Min Wang,[4] Lingling Qiao,[1,5] and Ya Cheng[1,3,4,5,7,8,9,*]

[1]*State Key Laboratory of High Field Laser Physics and CAS Center for Excellence in Ultra-Intense Laser Science, Shanghai Institute of Optics and Fine Mechanics (SIOM), Chinese Academy of Sciences (CAS), Shanghai 201800, China*

[2]*Research Center for Intelligent Sensing, Zhejiang Lab, Hangzhou 311100, China*

[3]*The Extreme Optoelectromechanics Laboratory (XXL), School of Physics and Electronic Science, East China Normal University, Shanghai 200241, China*

[4]*State Key Laboratory of Precision Spectroscopy, East China Normal University, Shanghai 200062, China*

[5]*Center of Materials Science and Optoelectronics Engineering, University of Chinese Academy of Sciences, Beijing 100049, China*

[6]*School of Physical Science and Technology, ShanghaiTech University, Shanghai 200031, China*

[7]*Collaborative Innovation Center of Extreme Optics, Shanxi University, Taiyuan 030006, China*

[8]*Collaborative Innovation Center of Light Manipulations and Applications, Shandong Normal University, Jinan 250358, China*

[9]*Shanghai Research Center for Quantum Sciences, Shanghai 201315, China*

[†]Electronic address: jintianlin@siom.ac.cn

[*]Electronic address: ya.cheng@siom.ac.cn

Date: Sept. 22, 2022

**Photonic-based low-phase-noise microwave generation with real-time frequency tuning is crucial for a broad spectrum of subjects, including next-generation wireless communications, radar, metrology, and modern instrumentation. Here, for the first time to the best of our knowledge, narrow-bandwidth dual-wavelength microlasers are generated from nearly degenerate polygon modes in a high-Q active lithium niobate microdisk. The high-Q polygon modes formation with independently controllable resonant wavelengths and free spectral ranges is enabled by the weak perturbation of the whispering gallery microdisk resonators using a tapered fiber. The stable beating signal confirms the low phase-noise achieved in the tunable laser. Owing to the high spatial overlap factors between the two nearly degenerate lasing modes as well as that between the two lasing modes and the pump mode, gain competition between the two modes is suppressed, leading to stable dual-wavelength laser generation and in turn the low noise microwave source. The measured microwave signal shows a linewidth of ~6.87 kHz, a phase noise of ~-123 dBc/Hz, and an electro-optic tuning efficiency of -1.66 MHz/V.**

A frequency-tunable low-phase-noise microwave source is highly desirable for many applications such as radar, wireless communications, software-defined ratio, and modern instrumentation.[1-3] Conventional microwave signals are generated using complicated and costly electronic circuits, severely suffering from low bandwidth and high-loss propagation in the electrical lines. In contrast, the generation of optically carried microwaves based on optical heterodyning offers superior solutions with broad bandwidth and low-loss microwave signal distribution. To this end, compact on-chip integrated microlasers have been developed for the generation of high-performance microwave signals featuring low phase-noise performance (e.g., linewidth on the order of magnitude of 10 MHz), compact device footprint, low energy consumption, ease of tunability, and broad wavelength coverage range.[4-9] Among the demonstrated dual-wavelength coherent light sources, on-chip dual-mode microlasers generated in common microcavities have attracted significant attention as the two laser modes in a single cavity are naturally phase locked without the need of external phase stabilizers. The miniaturized microwave synthesizers also provide indispensable high performance microwave sources for monolithically integrated fully functional microwave photonics.

In the past, stable and frequency-tunable microwave signals from several to hundreds of gigahertz have been successfully synthesized from dual-mode microlasers in deformed microcavities.[7-9] To avoid longitudinal mode competition in the cavity, transverse polygon modes are generated by introducing large cavities deformation which breaks the cavities azimuthal symmetry.[9] Yet it remains a challenge to achieve high-Q modes in the strongly deformed microcavities, resulting in the relatively broad linewidth of the synthesized microwave. In addition, rapid tuning of the microwave signals is highly desired but still missing due to the lacking of the suitable electro-optic substrates such as lithium niobate for microcavity fabrication.

Recently, thin-film lithium niobate (TFLN) has shown great potential for integrated photonic devices due to the excellent optical properties including large electro-optic coefficient and nonlinear optical coefficient, broad transparency window, and piezo-electric effect.[10-24] Using the TFLN platform, integrated electro-optic modulators, nonlinear optical frequency converters, compact frequency comb, as well as on-chip microlasers and optical amplifiers have been demonstrated, showing unparalleled device performances in terms of optical loss, speed, nonlinear wavelength conversion efficiency, optical gain coefficient, and tunability.[24] Recently, we have reported a narrow-linewidth microlaser in a weakly perturbed erbium ion-doped microdisk employing an ultra-high Q polygon mode.[25,26] Since the polygon modes are coherently synthesized with combination of the whispering gallery modes (WGMs) in a circular microdisk evanescently coupled with a tapered fiber, loaded Q factors as high as $10^7$ are achieved due to the weak perturbation from the tapered fiber which leads to slight degradation of the high Q factors of the WGMs. The unique advantage gives rise to the ultra-narrow linewidth single-mode lasing.[25,26] One may expect to generate stable dual-mode microlasers from two longitudinal polygon modes by merely increasing the cavity size (i.e., reducing the mode wavelength spacing), which is not possible because of the longitudinal mode competition.[9]

In this Letter, we demonstrate the generation of frequency-tunable narrow-linewidth microwave signals using dual-wavelength microdisk lasers fabricated on erbium ion-doped lithium niobate. The two laser modes are characterized to be nearly degenerated polygon dual modes with a small wavelength spacing in the same microdisk. The beating of the two lasers generates the low-phase-noise microwave signal. We also reveal the underlying physics using numerical simulation.

In our experiment, the active microcavities were fabricated on a TFLN wafer using the photolithography assisted chemo-mechanical etching technique.[26,27] The TFLN wafer was prepared by bonding an X-cut erbium ion-doped with concentration of 0.1 mol% TFLN with 710 nm thickness onto a holder wafer with 2 μm thickness $SiO_2$ and 500 μm thickness lithium niobate. The detailed fabrication process of the microdisk integrated with the microelectrodes is described in the **Methods** section, and the diameters of the fabricated LN microdisks supported by a silica pillar and microelectrodes on the top surface of the microdisks are 106.33 μm and 15 μm, respectively. The scanning electron microscope (SEM) image of the microdisk is shown in the inset of Fig. 1, indicating an ultrasmooth surface.

Figure 1 schematically shows the experimental setup for generating the frequency-tunable low-phase-noise microwave signals using dual-wavelength microdisk lasers fabricated on erbium ion-doped lithium niobate. The pump laser was provided by a narrow-linewidth tunable laser (Model TLB-6719, New Focus Inc.) together with a variable optical attenuator (VOA), and coupled into the microdisk through a tapered fiber with a diameter of 2 μm. The relative position of the tapered fiber can be accurately adjusted and monitored by a 3D piezo-electric stage with a resolution of 20 nm and a real-time optical microscope imaging system, respectively. The detail of the optical microscope imaging system is described in the **Methods** section. An in-line fiber polarization controller can adjust the polarization state of the pump laser. Experimentally, to efficiently excite the transverse-magnetic polarized polygon modes, the relative position between the tapered fiber and the microdisk center is 47 μm. The lasing experimental was carried out at room temperature, and a thermoelectric cooler (TEC) was placed underneath the microdisk chip to overcome environmental temperature fluctuations with a TEC driver of 2 mK temperature stability. The generated laser signals were separated by a beam splitter with power ratios of 20/80. The minority part was sent into an optical spectral analyzer (OSA) for spectral analysis. Meanwhile, the majority part was first filtered with a long-pass filter (Model FELH 1100, Thorlabs Inc.), and then amplified by an erbium-ytterbium-doped fiber amplifier (EDFA), and finally detected by a fast photodetector (bandwidth 30 GHz) to generate the beating signal. The beating signal and its phase noise were analyzed and recorded by a real-time electric spectral analyzer (ESA) with a bandwidth of 25 GHz.

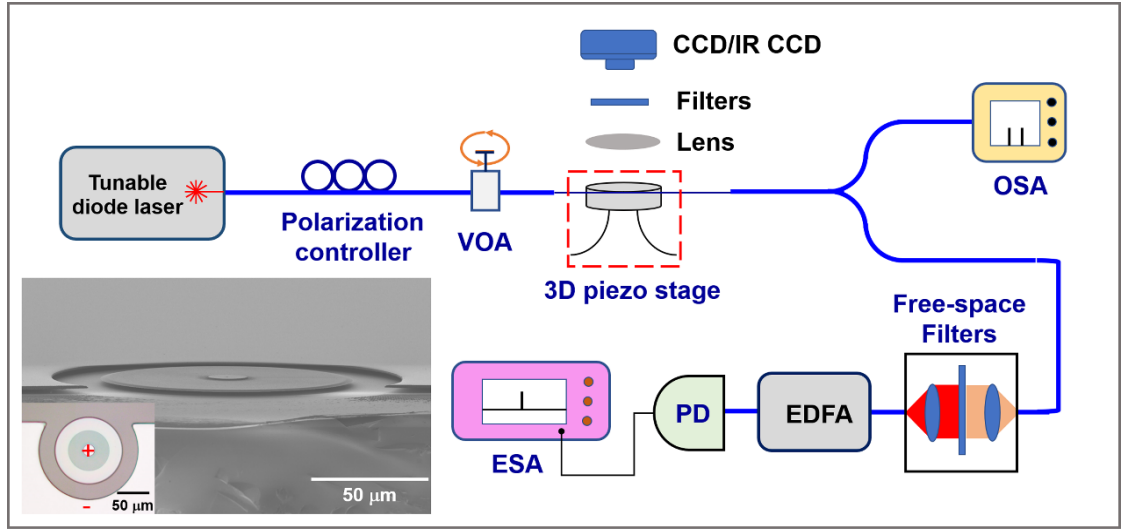

FIG. 1: **The experimental setup** for generation of microwave signal from dual-wavelength lasing in a single microdisk resonator. **Inset**: SEM image and optical micrograph of the microdisk integrated with microelectrodes.

The dual-wavelength lasing signals with a wavelength interval of 8 pm were observed when the pump wavelength and power are tuned to 970.02 nm and higher than 80 μW, respectively, as demonstrated in Fig. 2(a). There is an error of the measured wavelength interval, resulting from the resolution of the OSA with 20 pm. Correspondingly, the up-conversion fluorescence, pump and lasing modes emitted from the microdisk were captured by the optical microscope imaging system, as shown in Fig. 2(b). All the images feature octagon modes, displaying polygon modes with similar spatial distributions. When the pump wavelength was tuned to be 970.02 nm, a wavelength of the localized polygon mode, only the nearly degenerate polygon modes within the optical gain of erbium ion would be excited because of the considerable overlap between the pump and lasing modes. Figure 2(c) gives laser output power recorded at different pump power, indicating a lasing threshold of ~ 80 μW. The dual-wavelength lasing signals with a wavelength interval of 70 pm were observed when the positive position of the tapered fiber and pump wavelength are tuned to a relatively outer position of 0.5 μm and 969.82 nm, respectively, as shown in Fig. 2(d). We find the polygon patterns in Fig. 2(d) give a more approximate octagon shape. Thus, the wavelength interval of dual-wavelength lasing signals can be adjusted utilizing the perturbation induced by the tapered fiber coupling.

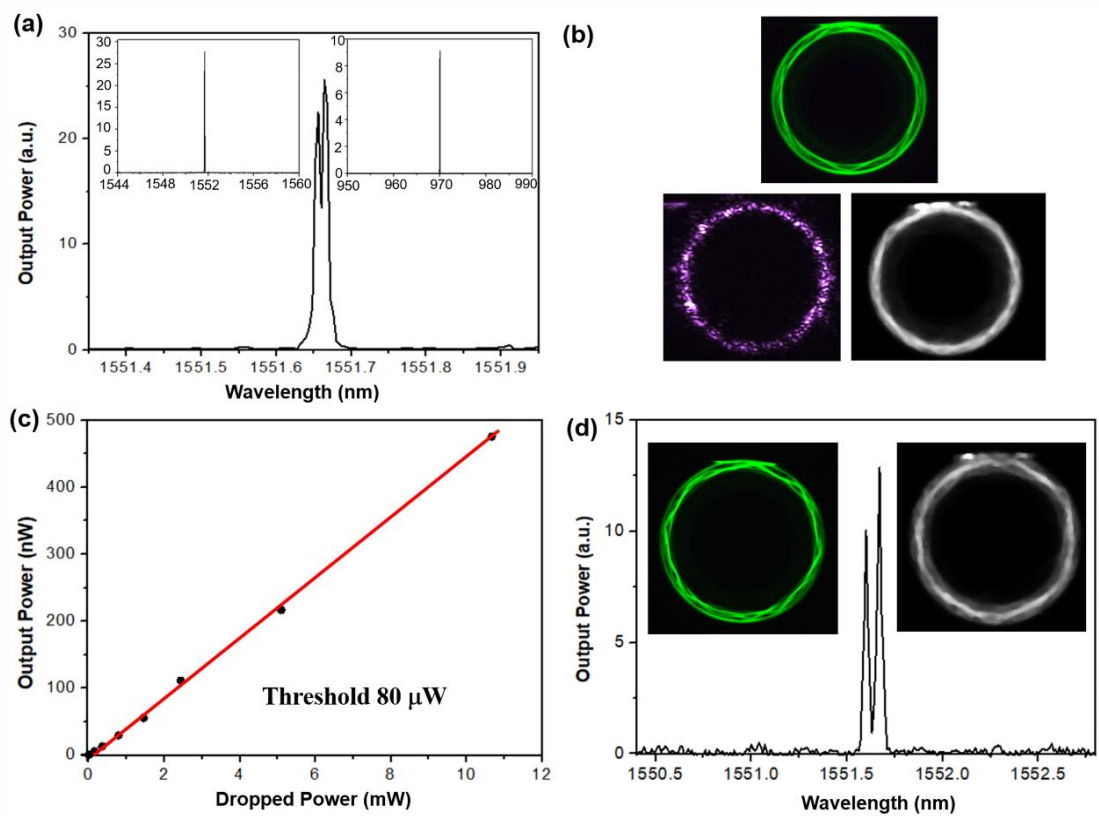

FIG. 2: **Dual-wavelength microdisk lasers. a.** The spectrum of the microlaser with fine structure. **Inset**: The spectrum of the microlaser shown in a wide wavelength range (left), indicating only one set of dual-wavelength signal, and the spectrum of the pump light (right). **b.** The optical microscope images of spatial intensity distributions of the up-conversion fluorescence (top middle), pump mode (bottom left), and lasing modes (bottom right). **c.** Output power dependence of the pump power. **d.** The other spectrum of dual-wavelength microlasers with the varied coupled position. **Inset**: The optical microscope images of spatial intensity distributions of the up-conversion fluorescence (left) and lasing modes (right).

The mode structures of the microdisk around 970 nm and 1550 nm waveband were characterized by tunable diode lasers (Models: TLB-6719 & TLB-6728, New Focus, Inc.) with an output power of 5 μW, at the same coupled position as the dual-wavelength lasing experiment of the wavelength interval of 8 pm. Figure 3(a) and 3(c) show the transmission spectra measured by a photodetector with 125 MHz bandwidth around the pump and lasing wavelengths, respectively. The Q factor of the mode at the pump wavelength is estimated to be

$3.5\times10^6$ by Lorentz fitting of the resonant dip, as plotted in Fig. 3(b). Figure 3(d) demonstrates the magnified spectra indicated by a dotted box in Fig. 3(c). Coincident dual-dip structures with a wavelength interval of 9.9 pm can be observed, and the Lorentz fitting curves indicate a pair of lasing modes with Q factors of $8.7\times10^6$ and $9.0\times10^6$, respectively.

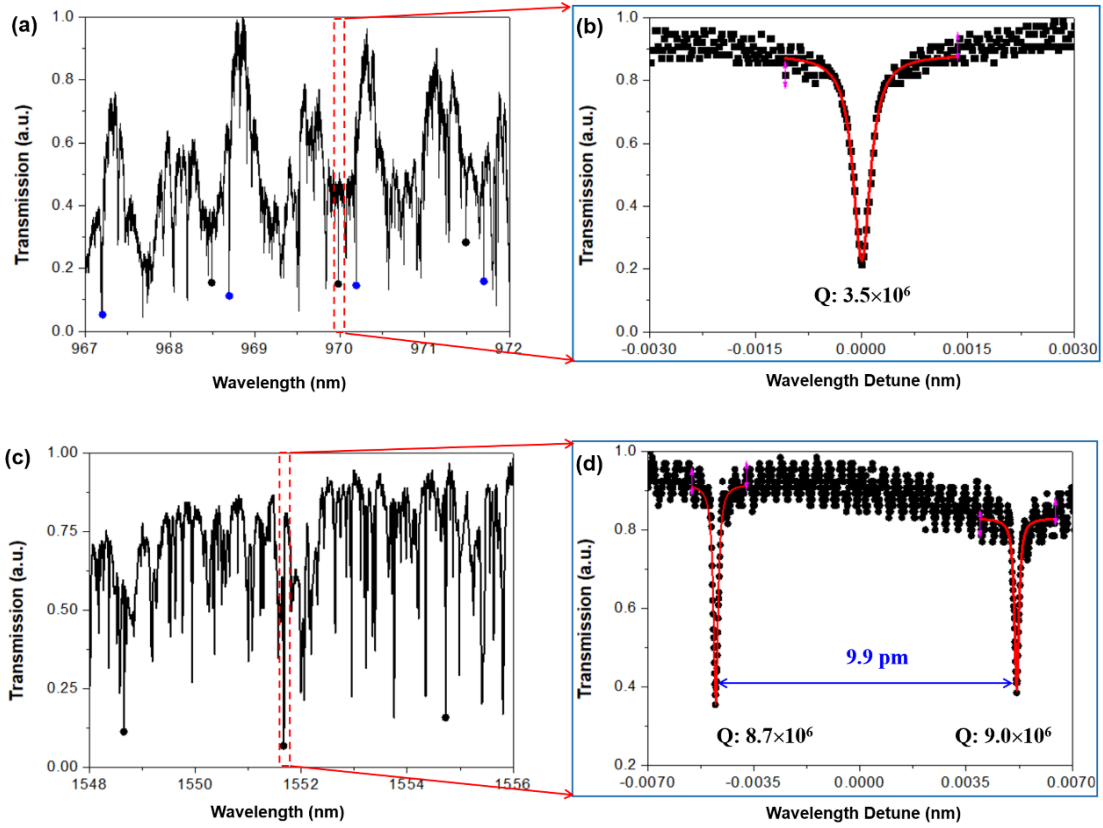

FIG. 3: **Transmission spectra of the tapered fiber coupled with the microdisk. a.** Transmission spectrum around the pump wavelength, where the dips labeled with black and blue dots are the families of the polygon and fundamental modes, respectively. **b.** The Q factor of the pump mode. **c.** Transmission spectrum around the lasing wavelengths. **d.** The Q factors of the lasing modes.

The majority part of the output power of the microlaser around 1551. 66 nm wavelength was amplified to 0.047 mW to synthesize microwave signal with the fast photodetector by beating. The quality of the generated beat note signal was analyzed by the ESA with a resolution of 80 Hz, indicating a near Lorentzian-shaped power signal with a center frequency of 1.23 GHz is generated, as exhibited in Fig. 4(a). Since the frequency/phase noise consists of white noise and $1/f$ noise,[4] where $f$ is the frequency, the spectral width of the microwave signal was estimated to be 20 dB down from the peak and the Lorentzian 3 dB linewidth was extracted to be 6870 Hz. Figure 4(b) records the phase noise of the microwave carrier. When the frequency offset is more than 110 MHz, corresponding to a white-phase-noise floor of the carrier signal of -123 dBc/Hz, the phase noise is almost flat, indicating a highly stable microwave signal and dual-wavelength microlasers. Figure 4(c) shows the electro-optic tuning of the microwave signal, here linear electro-optic coefficient $r_{22}$ is employed for tuning the wavelength interval of the nearly degenerate modes.[26] When the externally applied direct-current bias is tuned from -300 V to 300 V, the microwave signal is red-shifted from 1.73 to 0.73 GHz, showing the tuning efficiency of -1.66 MHz/V. The dual-wavelength lasing signal with a wavelength interval of 70 pm is also utilized to synthesize the microwave signal, and correspondingly, the microwave signal is located at 8.72 GHz, as shown in Fig. 4(d). Thus, a broad frequency range tuning of the generated microwave frequencies can be directly realized by adjusting the perturbation location of the microdisk.

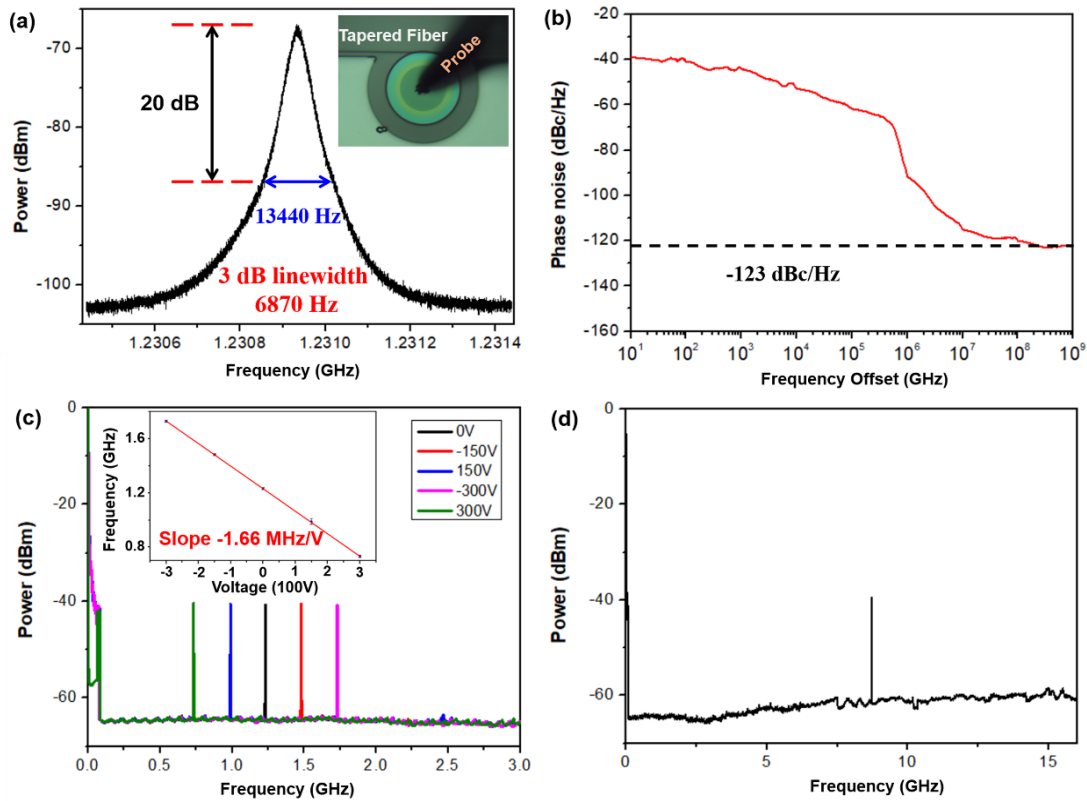

FIG. 4: **Microwave signal synthesis and tuning. a.** The intensity noise of the microwave signal. **Inset**: optical micrograph of the microdisk coupled with the tapered fiber, when the Cr/Au probe was contacted with the microelectrode. **b.** The phase noise of the microwave carrier. **c.** The tuning of the microwave signal. **Inset**: the tuning efficiency. **d**. The microwave signal synthesized from the other dual-wavelength microlaser with large wavelength interval.

The underlying physics behind the formation of nearly degenerate polygon modes with high-order excited states[21] is further theoretically analyzed using a 2D simulation. In our simulation, perturbation is introduced into the circular microcavity through coupling with the tapered fiber by adjusting the disk-taper distance,[25,26] other than cavity deformation.[28-31] With the help of the tapered fiber coupling, nearly degenerate polygon modes are formed with the recombination of inherent nearly degenerate eigenmodes of the thin circular microdisk. The intensity profiles of the first and the second excited state of the polygon modes at the lasing wavelengths are demonstrated in Figs. 5(a) and (b), exhibiting octagon patterns with $\pi$-phase difference. Here, the polygon mode exhibits octagon pattern. It is noteworthy that the octagon pattern appears slightly incomplete because of the weak perturbation of the fiber. The calculated 10 pm wavelength interval of the two nearly degenerate modes agrees well with the measured result. The intensity profile of the pump mode is plotted in Fig. 5(c), showing a slightly incomplete octagon-shaped pattern as well. The spatial overlap factors $\Gamma = \frac{\langle I_1|I_2\rangle}{\sqrt{\langle I_1|I_1\rangle\langle I_2|I_2\rangle}}$ between the pump mode $I_1$ and the nearly-degenerate lasing modes $I_2$ are calculated. The corresponding overlap factors are 0.63 and 0.52 when the distributions of the integrand between the pump and lasing modes are calculated, as shown in Figs. 5(d) and (e), respectively. Remarkably, the dual-wavelength signals have almost the same gain due to the overlap factors being close. Moreover, both the distributions of the integrand calculated in Figs. 5(d) and (e) show a periodic wave-like structure but with a phase shift of $\pi$ along the periphery of the microdisk. The maximum overlaps of Fig. 5(d) always correspond to the zero points of Fig. 5(e), and vice versa. This characteristic provides an extremely useful advantage for generating the stable dual-wavelength laser output and in turn for synthesizing the low phase noise microwave because of the strong suppression of the longitudinal mode competition.

To conclude, we have demonstrated frequency-tunable narrow-linewidth microwave signals using dual-wavelength lasers from a single microdisk fabricated on erbium ion-doped lithium niobate thin film. Nearly degenerate high-Q polygon modes with $\pi$-phase difference were formed in the weakly perturbed circular microcavity, which triggers stable dual-wavelength lasing and the resulting stable microwave signal directly from the integrated platform. The linewidth of the highly coherent microwave signal is measured as narrow as 6.87 kHz, being at least three orders of magnitude narrower than the previously reported results from dual-wavelength integrated lasers. Such integrated microwave source is doomed to promote future advancements of microwave photonics and optical information processing.

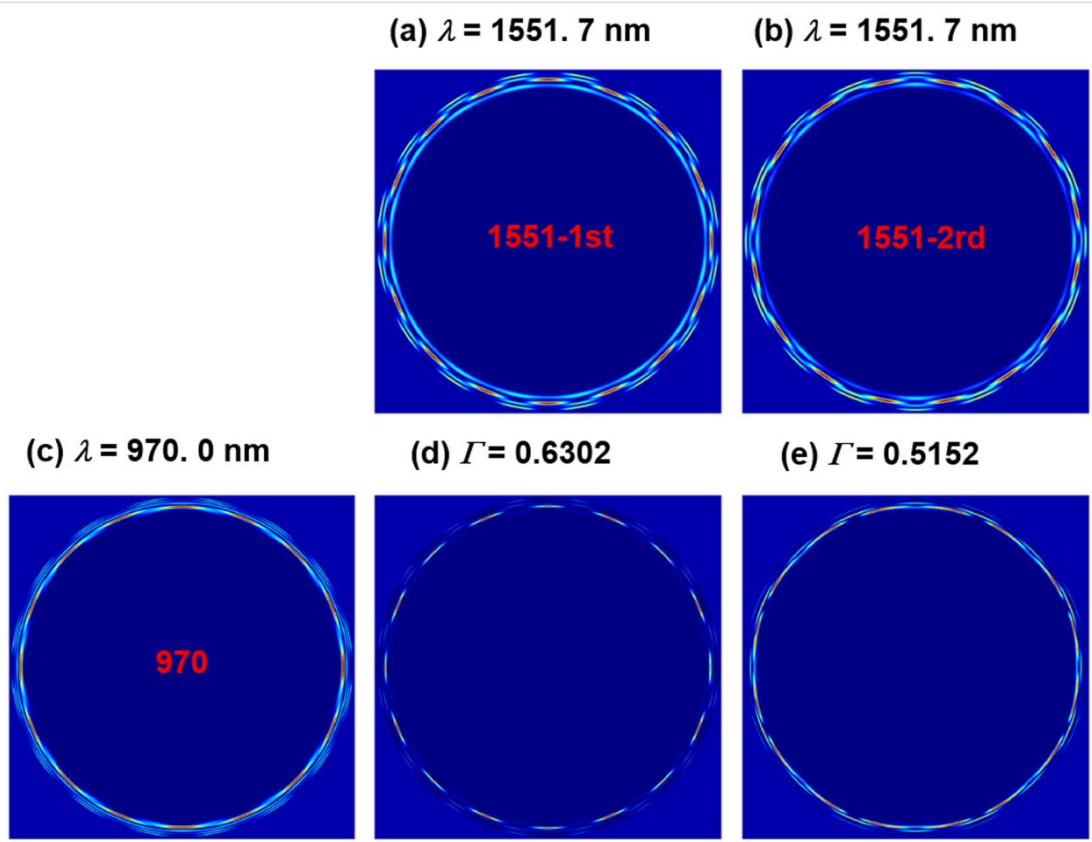

FIG. 5: **The intensity distributions** of (a) and (b) the nearly-degenerate polygon lasing modes and (c) the pump polygon mode. (d) and (e)The intensity distributions of the integrand between laser modes at the first and second excited states illustrated in the top row and the pump mode, respectively.

**Methods**

**Fabrication.** The microdisks integrated with microelectrodes are fabricated by photolithography assisted chemo-mechanical etching (PLACE) [19]. The fabrication process begins from the preparation of erbium ion doped LN thin film by ion slicing. The sample endures chromium (Cr) layer coating, hard mask patterning via femtosecond laser ablation, pattern transferring from the Cr hard mask to LN thin film via chemo-mechanical etching, and chemical wet etching to completely remove the Cr layer and partially remove the silica layer underneath the LN disk into the supporting pedestal. The details of the fabrication process can be found in Ref. [26,27].

**Measurement.** The optical microscope imaging system consists of an objective lens with a numerical aperture of 0.42, long-pass or short-pass filters, and a visible or infrared charge-coupled device (CCD). To capture the up-conversion fluorescence and pump signals, short-pass filter (Model FES 800, Thorlabs Inc.) and long-pass filter (Model: FELH 800, Thorlabs Inc.) are inserted before the visible CCD, respectively, and the lasing signals are captured by long-pass filter (Model FELH1100, Thorlabs Inc.) and the infrared CCD. The dual-wavelength lasing signals are coupled out of the microdisk by the fiber taper once the pump power reaches the lasing threshold.